\pdfoutput=1
\documentclass[doublecol,final]{epl2}

\usepackage{graphicx}
\usepackage{hyperref}
\usepackage{amsfonts}
\usepackage{amsmath}
\usepackage{listings}
\usepackage[T1]{fontenc}
\usepackage[utf8]{inputenc}
\usepackage[english]{babel}
\usepackage[above,below]{placeins}

\begin{document}

\title{Sampling eigenmodes in colloidal solids}
\author{A.C. Maggs \and M. Schindler}
\institute{Laboratoire PCT, Gulliver CNRS-ESPCI UMR 7083, 10 rue Vauquelin,75231 Paris Cedex 05, France}
\date{\today}%
\abstract{We study the properties of correlation matrices widely used
  in the characterisation of vibrational modes in colloidal
  materials. We show that the eigenvectors in the middle of the
  spectrum are strongly mixed, but that at both the top and the bottom
  of the spectrum it is possible to extract a good approximation to
  the true eigenmodes of an elastic system.}
\vol{109}\year{2015}\firstpage{48005}
\doi{10.1209/0295-5075/109/48005}

\maketitle

The excitation spectrum of crystalline but also disordered colloidal
solids has recently been studied in both two
\cite{maret,maret2,andrea,pen} and three \cite{ghosh,antinaPRL}
dimensions: Experiments typically image a thousand or so micron-sized
particles; from a video recording, computer analysis is used to extract
a matrix of correlated fluctuations. The hope is that the spectrum and
the eigenvectors of the correlation matrix can be used to deduce
interesting properties of the colloidal material
\cite{wyart,wyart2,wyart3}, including local modes and incipient soft
structures, or even three dimensional elastic properties
\cite{claire}. Most experimentalists work with the matrix
\begin{equation}
  C_{ij} = \langle \delta r_i \delta r_j \rangle = \frac{1}{ T} \sum_{t=1}^T  \delta r_i(t) \delta r_j(t)  \label{eq:corr}
\end{equation}
where $\delta r_i(t)$ denotes a transversion fluctuation (in $x$ and
$y$ when imaging along~$z$) of a particle at time~$t$.  For a system
of $N$~particles this matrix has dimensions~$2N\times 2N$. If the
particles are coupled with linear springs the correlation matrix can
be related to the interactions as follows
\begin{equation}
  C=\frac{1}{\beta A}
\end{equation}
where~$A$ is the dynamical matrix of the system -- at least in the
limit of large~$T$. Thus the eigenvectors of~$C$ and~$A$ should be
identical and there should be an inverse relationship between the
eigenvalues of the two matrices.  Even in hard sphere systems the mode
structure approximates this linear Ansatz. $\beta$ is the inverse
temperature.

\begin{figure}[tb]
  \centering
  \includegraphics[width=0.48\textwidth]{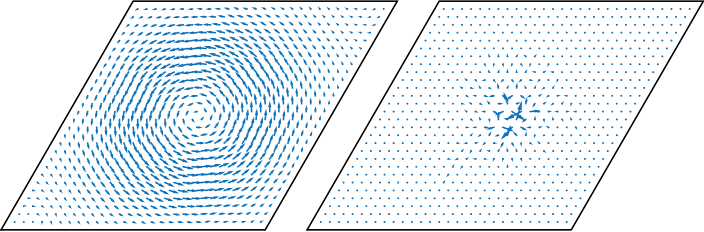}%
  \caption{Left: A low-energy mode for an elastic medium
    eq.~(\ref{eq:el}). Right: High-energy localised mode.  Hexagonal lattice
    with Dirichlet boundary conditions. $N=28^2$
    particles.}\label{fig:third}%
\end{figure}

In previous work we considered the question of projection of the modes
from three to two dimensions~\cite{claire,truncate}. In this paper we
consider the effect of observation statistics on the mode structure. It is
already well known~\cite{silke} that the use of a number of recordings~($T$)
which is smaller than the number of observed degrees of freedom~($2N$) leads to
a rank-deficient matrix~$C$ for which many eigenvalues
are zero. Even when $T>2N$ the theory of Marchenko and Pastur
\cite{pastur2} shows that there are large systematic (i.e.\
non-statistical) errors which appear in the spectrum. In fact the
observed spectrum is deterministically distorted as a function of
$z=T/(2N)$.

The rather remarkable results on the evolution of the spectrum of the
correlation matrix are not matched by a detailed theory of the
evolution of the eigenvectors; results such as those in ref.~\cite{jp} tell us about
some angular correlations but do not contain the all the information
needed by experimentalists to interpret typical data sets.  It seems
clear that statistical and systematic noise in the sum
in eq.~\eqref{eq:corr} will mix eigenmodes, in a manner which is familiar
from perturbation theory in quantum mechanics. The point of the
present paper is to quantify this mixing in order to give simple rules
of thumb as to how many modes can be trusted in a correlation
analysis.

Some authors give examples of eigenmodes extracted from the matrix~$C$,
often coming from different places in the spectrum -- for instance low-energy modes, high-energy modes or modes coming deep within the
spectrum, such as near a van Hove peak. In the elastic model that we
consider a typical low-energy mode is shown in Fig.~\ref{fig:third},
left. On the right of Fig.~\ref{fig:third} we see a high-energy,
localised mode of $A$ in a disordered elastic medium.  {Typically the
  mode is represented as a series of arrows, with amplitude  
  proportional to the component of the eigenvector at the particle
  position.}

Our main conclusion is that the bottom of the spectrum of~$A$,
including modes such as that depicted Fig.~\ref{fig:third},left is
reproduced rather easily on diagonalisation of the correlation
matrix. The top of the spectrum depends on the model considered: When
we consider the elastic vibrations of a disordered system with strong
localisation the top of the spectrum also converges for moderate
numbers of samples (though less well than at the bottom). In all cases
{\it the middle of the spectrum leads to mixing of an extensive number
  of eigenvectors}, so that little information on the true mode can be
observed using a correlation analysis. This is particularly the case
near van Hove singularities which seem to strongly favour mode mixing.
However, without disorder even the top of the spectrum is badly
reproduced; mode reconstruction clearly contains components which
are more model dependent than the remarkable Marchenko-Pastur result
which depends only on the density of states of the original system.

The experimentalist should also worry about the effect of the finite
auto-correlation time for any experimental system. This was studied
for the spectrum in a number of papers \cite{matrix1,pole}, but no
results are available for the effect of correlations on modes. Thus,
we conclude with a short study on the influence of finite relaxation rates on
the observed mode structure and show that the mode structure is
remarkably stable even in the presence of slowly relaxing modes.

\section{Elastic model}
In this paper we study the modes of a two-dimensional solid, firstly
because data in published experiments is recorded from two-dimensional
slices but also because two-dimensional elastic systems contain few
enough degrees of freedom so that we can use direct matrix solvers to
study the mode structure. The study of a three dimensional medium
would require more sophisticated iterative algorithms.

We work with a network of central springs with the energy
\begin{equation}
  U=\frac{1}{2}\sum_{ij} K_{ij}(r_{ij} -1)^2, \label{eq:el}
\end{equation}
where~$K_{ij}$ is the spring constant between particles $i$ and~$j$;
$r_{ij}$ is the separation between the particles. We either take the
spring constants as equal, in order to study a crystalline
material, or we take a model of site disorder where each particle is
characterised by a random stiffness~$k_i$. We set the bond constant
$K_{ij}^{-1} = k_i^{-1} + k_j^{-1}.$ We use a hexagonal lattice which
generates an isotropic elastic system obeying the Cauchy relation
between the shear and compression modulus~\cite{born98}.

From the energy we generate the matrix
\begin{equation}
  A_{ij} = \frac{\partial^2 U}{\partial r_i \partial r_j}
\end{equation} of second derivatives as
well as the Cholesky factorisation of~$A$ which is used to generate
the correlation matrix according to the Wishart
distribution for $T$~samples~\cite{wishart}. This corresponds to generating the
correlation matrix, $C$, as an average of $T$~statistically
independent samples. It ignores the possibly slow relaxation of modes
in a true experimental sample.

\begin{figure}[tb]
  \centering
  \includegraphics[width=0.45\textwidth]{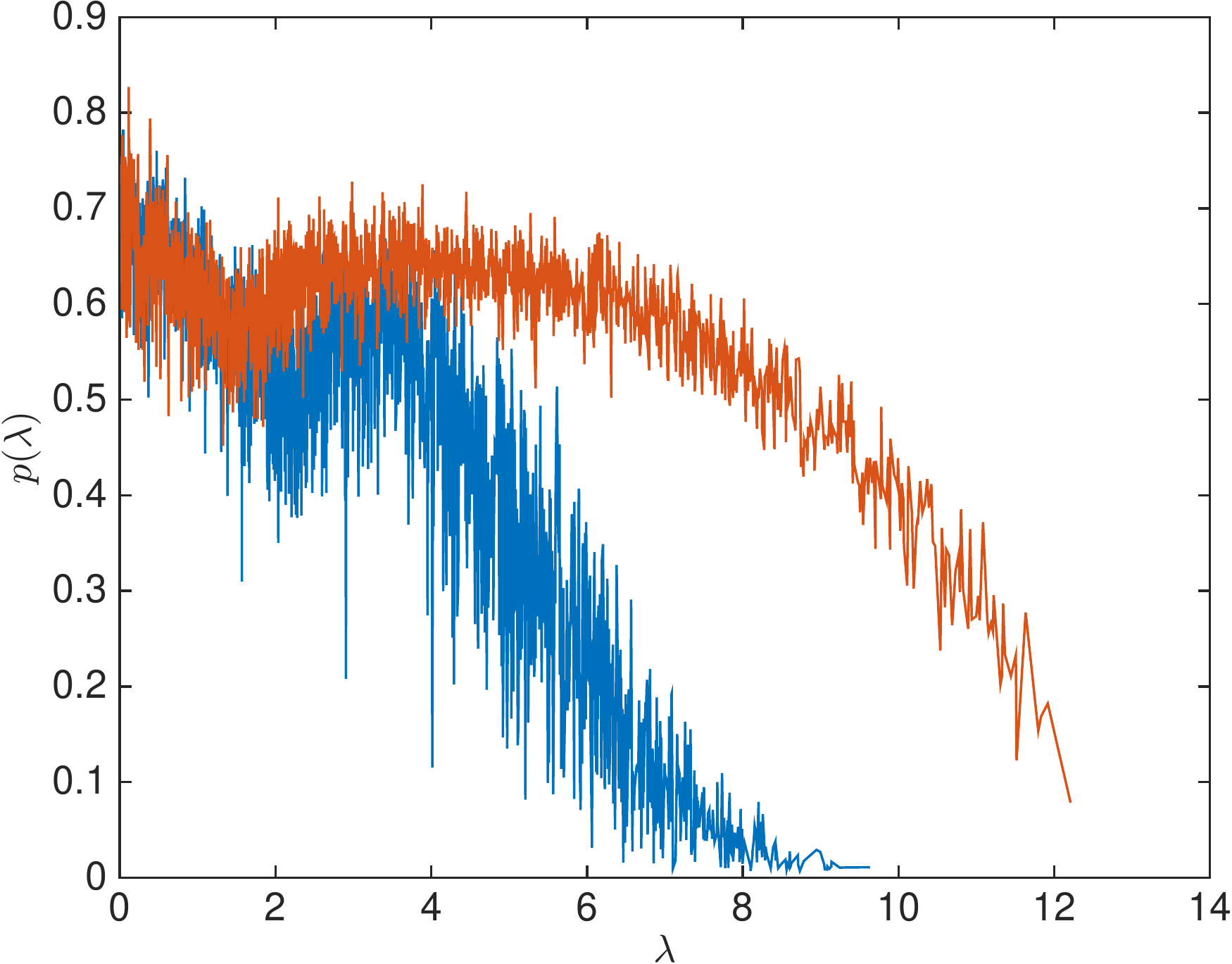}%
  \caption{ Participation ratios in a elastic medium with disordered
    elastic constants: from the dynamical matrix~$A$ (lower curve,
    blue), and from the diagonalised correlation matrix~$C^{-1}$
    (upper curve, red).  We see that the system is characterised by a
    number of high-energy, localised states. The sampled system
    reproduces this fact rather badly except at very top of the
    spectrum. $N=28^2$, $z=10$. Abscissa $\lambda$ for $A$
    and~$C^{-1}$. Note the different upper limits in the spectrum due
    to the Marchenko-Pastur law.}\label{fig:particip}%
\end{figure}

We generate a disordered medium with a strength of disorder which is
tuned so that the nature of the modes is qualitatively similar to that
observed in experimental systems~\cite{pen}. In particular we consider
in Fig.~\ref{fig:particip}
the participation ratio,
\begin{equation}
  \frac{1}{p(\lambda)} = N \sum_i v_{i \lambda}^4
\end{equation}
where $v_{i\lambda}$ is the $i$'th component of the normalized
eigenvector with energy~$\lambda$. In Fig.~\ref{fig:particip} we plot
$p(\lambda)$ as a function of~$\lambda$. $p$ measures, approximately,
the proportion of sites over which a mode is localised. For extended
modes $p$~is~$\mathcal{O}(1)$. For a mode which excites a single site,
$p$~is $\mathcal{O}(1/N)$. For an ordered system all modes are extended, however on
adding disorder we see that high-energy eigenmodes localize to just
a few sites, Fig.~\ref{fig:third}, right.  For the sampling $z=10$
used in Fig.~\ref{fig:particip} many modes of the matrix $C$ are
qualitatively different from those contained in $A$ for large~$\lambda$. In
particular the modes are often too extended.


\section{Characterizing the modes}

\begin{figure}[tb]
  \centering
  \includegraphics[width=0.45\textwidth]{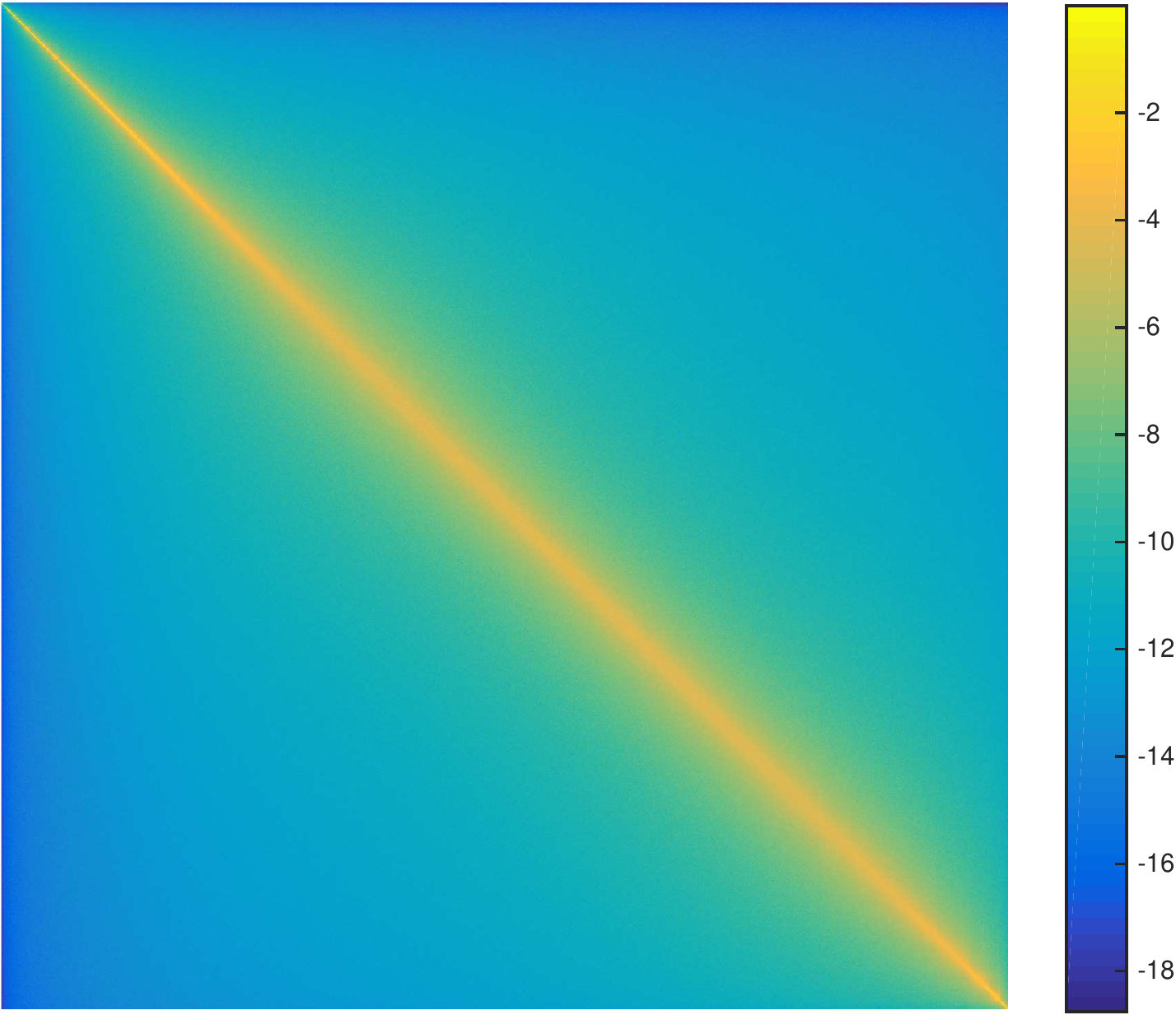}%
  \caption{ Visualisation of $W$ on a logarithmic scale in order to
    see the large number of elements with small amplitude.  The matrix
    is strongly diagonally dominant giving a bright central band.
    Disordered elastic medium, $N=28^2$, $z=60$ averaged over $20$~realisations
    of the disordered matrix~$A$. The diagonal dominance
    implies that it is above all modes with energies which are close
    in $A$ which are confused in the matrix~$C$.} \label{fig:overlap}%
  \centering
  \includegraphics[width=0.45\textwidth]{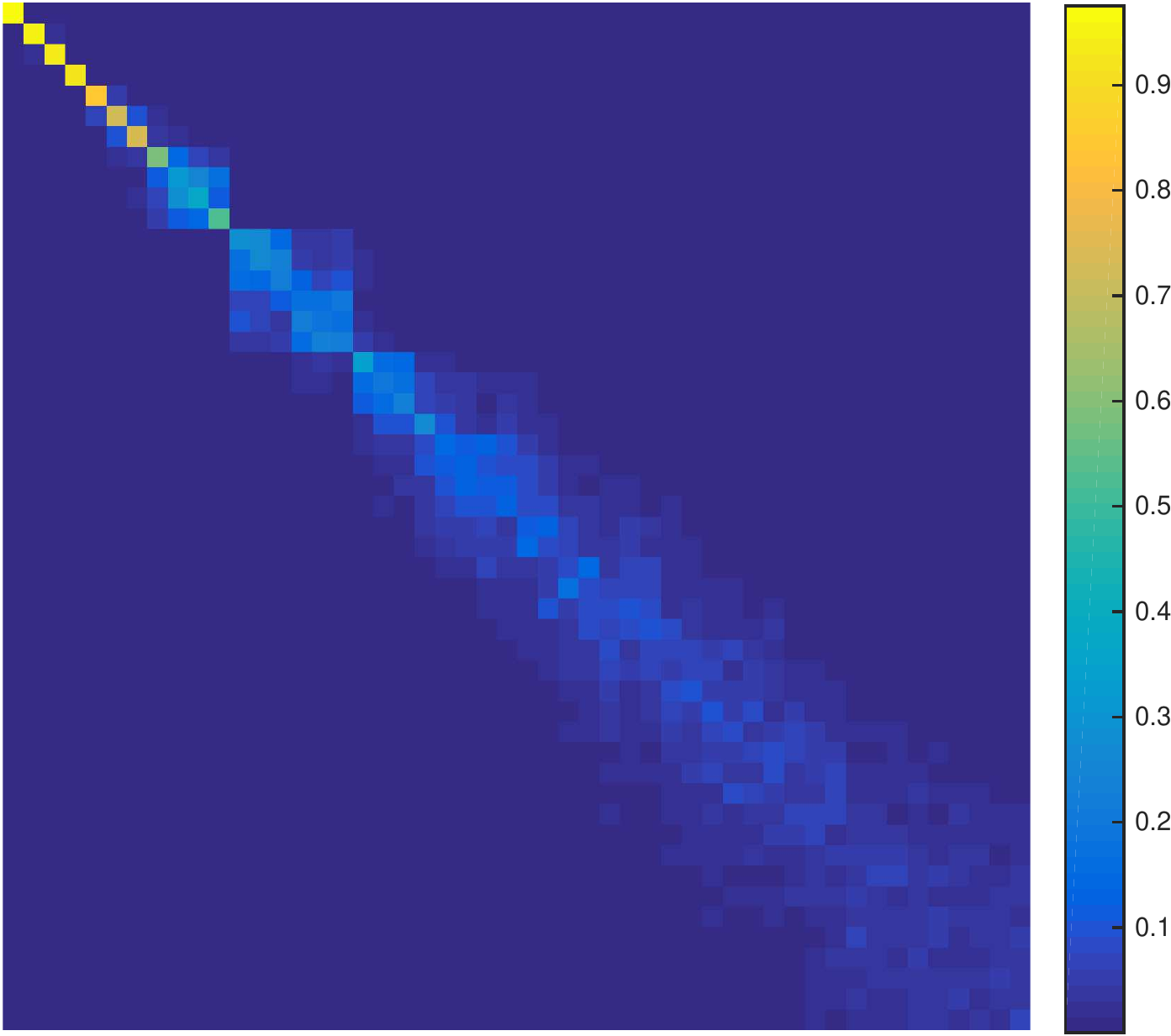}%
  \caption{Visualisation of the matrix~$W$ (linear scale)
    zoomed to the lowest energy states of~$A$. $N=28^2$. Here
    $C$ is recorded with very poor statistics so that~$z=0.75$.
    Despite this we see that several low-energy states of~$A$ are well
    reproduced in the matrix~$C$. Beyond the 7'th mode the amplitude
    is spread between several modes, which are however still close to
    the diagonal.}\label{fig:overlap2}%
\end{figure}

The main tool that we will use to characterise the similitude of modes
sampled experimentally and deduced from the original elastic system is
the set of overlaps
\begin{equation}
  W_{ij} =  (u_i\cdot v_j)^2
\end{equation}
Where $v_j$ is the $j$'th eigenvector of~$A$ and $u_i$~is an eigenvector
of~$C$, which form a matrix with indices describing the modes of $C$
and~$A$. We always sort the modes by increasing eigenvalues for~$A$, and
by decreasing eigenvalues for~$C$. In the case of perfect statistics the matrix
$W$ converges to the identity\footnote{There is a possible exception
  in a crystal where symmetry-related modes can mix.}. We also note
that each row and column of~$W$ sums to unity. We use deviations of
$W$ from the identity to quantify non-convergence of the experimental
eigenmodes to their final limit. In Fig.~\ref{fig:overlap} we plot $W$
on a logarithmic scale using a colour code to express the amplitude of
each element. We are firstly struck by the diagonal domination of the
matrix. The bright stripe indicates that modes mostly mix with other
modes with similar energies -- even if there is also a broad and
diffuse background. We also notice that the band is narrower in the
top-left and bottom-right corners which correspond to the bottom and the top
of the spectrum of~$A$. $W$~is more strongly diagonal for these modes
and thus the eigenvectors of~$C$ are close to eigenvectors of~$A$.

We confirm this point by plotting on a linear scale the top left
corner of the matrix~$W$ corresponding to the lowest-energy states of
$A$. In Fig.~\ref{fig:overlap2} the strong diagonal for the first
modes confirms that the lowest modes in the system are very well
reproduced in the correlation matrix. It is only on going higher in
the spectrum that we see the broadening which indicates that each
eigenvector of~$C$ is described by several eigenmodes of~$A$. We find
that when studying systems with $\mathcal{O}(1000)$ particles even when the
system is sampled with $z=0.75$ the very first mode is rather well
represented in~$C$, even though $C$~is a highly defective matrix.

\begin{figure}[tb]
  \centering
  \includegraphics[width=0.45\textwidth]{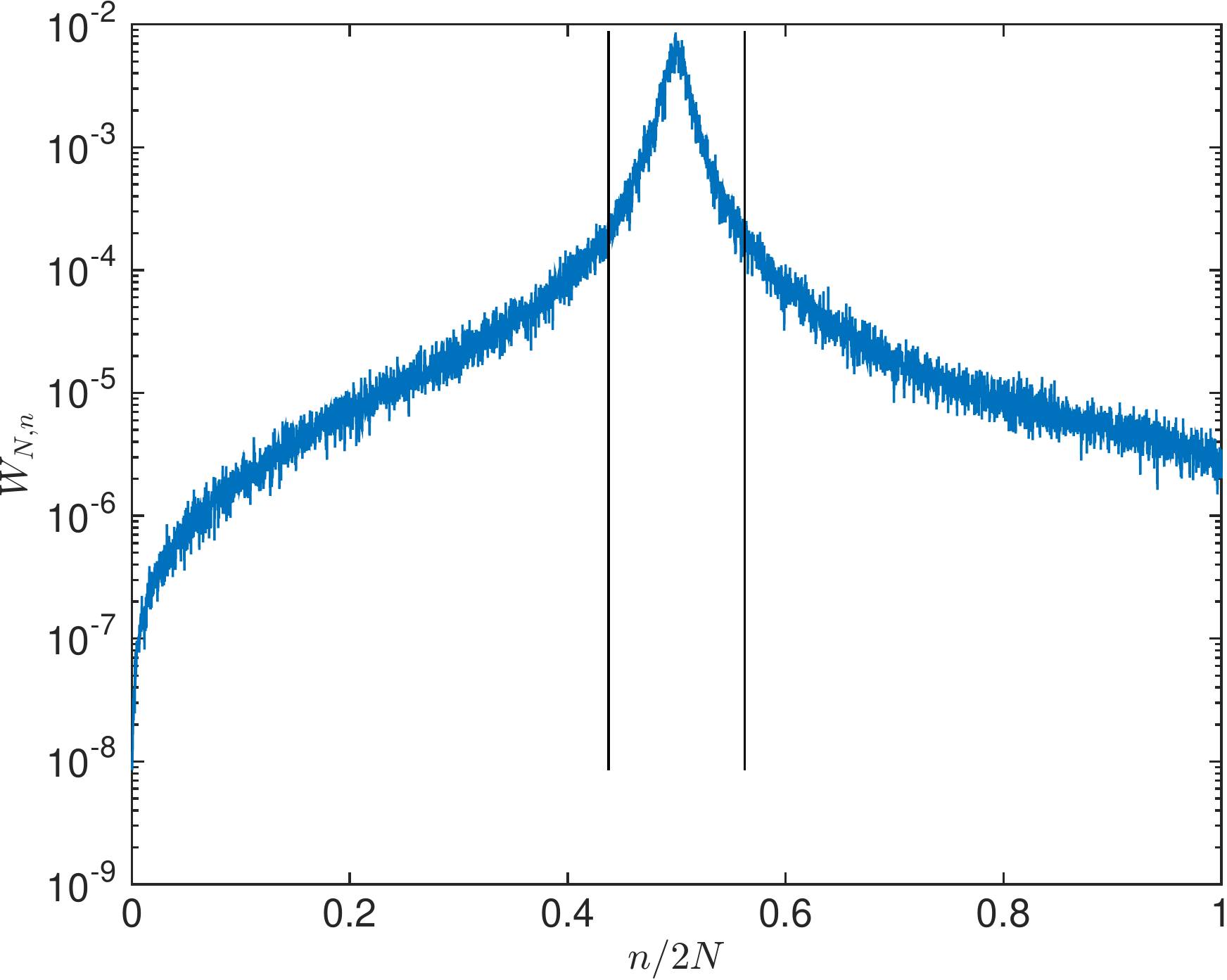}%
  \caption{A cut through the matrix $W$, corresponding to the central
    row of the matrix~$C$. A strong central peak is superposed upon a
    broader background, $N=48^2$, $z=60$. Note that the largest
    element of~$W$ in this slice has an amplitude close to $10^{-2}$
    so that hundreds of modes from~$A$ are needed to describe the bulk
    modes of~$C$. Vertical lines denote the 90\% band
    width~$b$.}\label{fig:middle}%
\end{figure}%
\section{How does mode mixing scale with $N$ and $z$?}
We now examine a row in the middle of the matrix $W$,
and plot the amplitude in a log-linear
scale in Fig.~\ref{fig:middle}. We see a sharp central peak, superposed on a broader
background. We tried to characterise the width of the central peak by
using moments of the distribution, however the result was
unsatisfactory due to the background in the figure. We chose an
alternative method of characterising the signal which was to take the
band-width, $b$, which contains 90\% of the amplitude. This measure is
much more robust to a broad outlying signal and is used to
characterise the spreading of modes for the rest of this paper.

\begin{figure}[tb]
  \centering
  \includegraphics[width=0.45\textwidth]{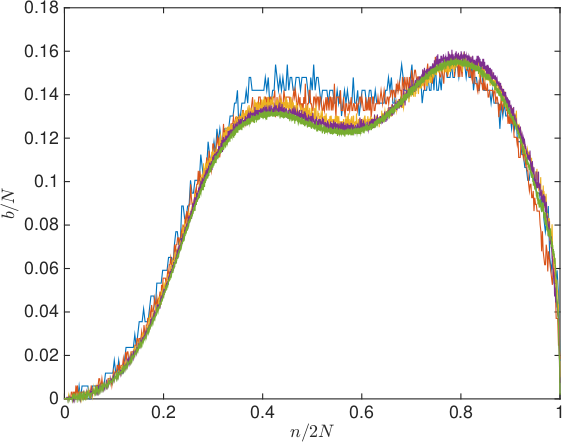}%
  \caption{90\% bandwidth, $b$ of $W$ for several different systems
    sizes, measured for $z=60$, $N=38^2$, $N=18^2$,
    $N=38^2$, $N=58^2$. Plotted from top left to bottom right of the
    matrix. The band width increases quadratically near the top left
    of the matrix, corresponding to the lowest-energy modes of~$A$. In
    the centre of the matrix there is a broad range of modes where an
    extensive fraction of eigenmodes are mixed together. For the
    highest-energy modes of~$A$ the matrix $C$ again gives a good
    representation of the mode structure for a very small number of
    modes. Disordered system.}\label{fig:sizes}
  \centering
  \includegraphics[width=0.45\textwidth]{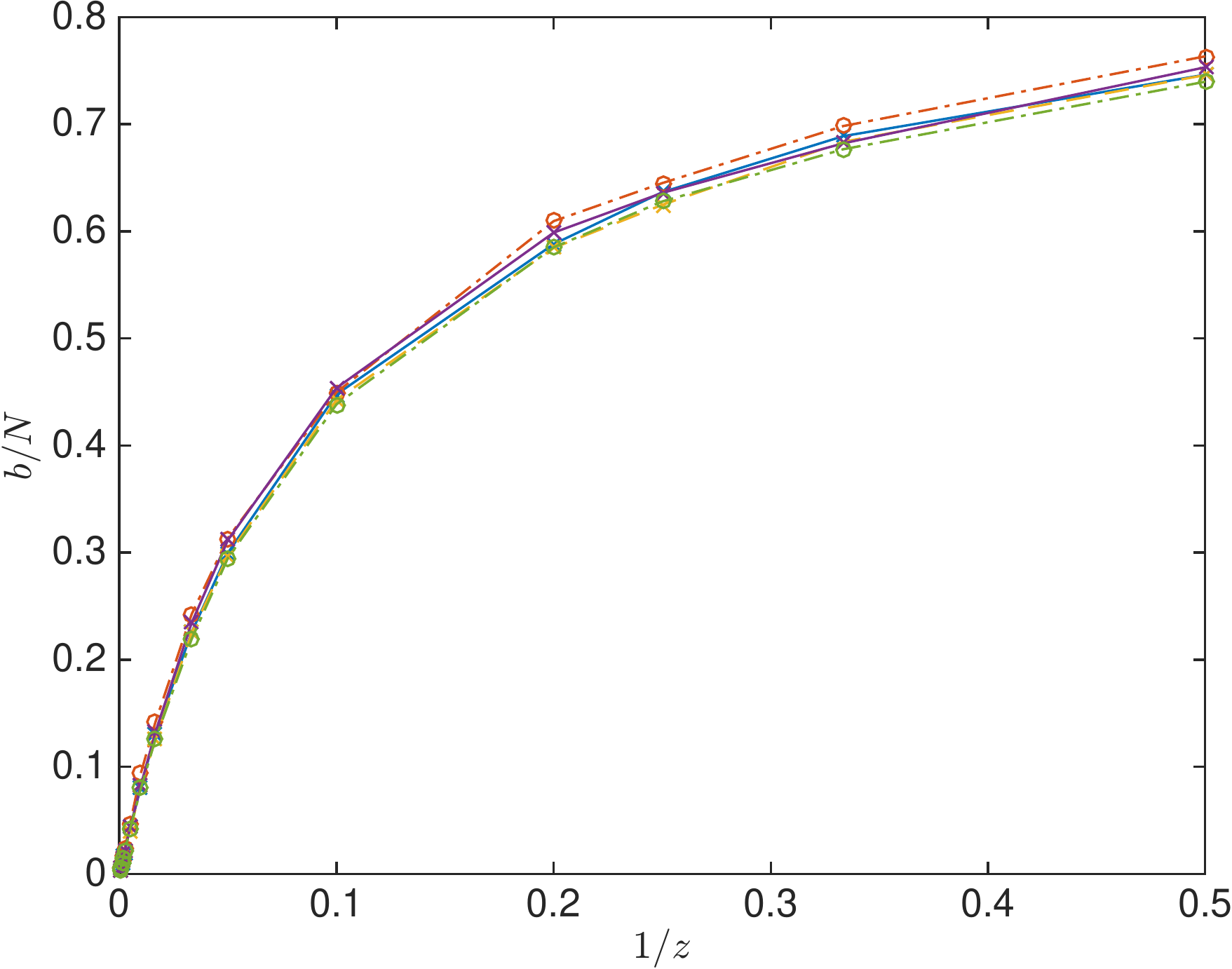}%
  \caption{90\% bandwidth of the middle mode of~$W$ as a function of the inverse
    sampling, $1/{z}$. Data for different system sizes fall onto a single master
    curve. $N=18^2$, $N=28^2$, $N=38^2$, $N=58^2$. Disordered system.}\label{fig:band}%
\end{figure}

In Fig.~\ref{fig:sizes} we choose several different sample sizes and
plot the 90\% width, $b$ scaled by the number of particles as a
function of the rank of the mode in the spectrum,~$n$. Each
correlation matrix was recorded with $z=60$ corresponding to a high-statistics
experiment. We see that all system sizes behave in a
similar manner. For both high and low energies the band-width of the
matrix is small, but for most interior modes in $C$ the band width is
$\mathcal{O}(0.16 N)$.  Thus a single mode in $C$ is actually a
mixture of an extensive number of modes in~$A$. In Fig.~\ref{fig:band}
we consider a single system size and plot $b$ as a function
of~$1/{z}$. The curve comes to the origin linearly: It seems clear
that extracting accurate eigenmodes in the middle of the spectrum is a
very difficult task requiring very large values of $z$ if the only
information available is the correlation matrix~$C$.

We also performed a similar study on the matrix~$C$ for an ordered
elastic medium in which all the spring constants are identical. The
conclusions for the low-lying and middle modes of~$A$ are very
similar. However curves such as Fig.~\ref{fig:sizes} are rather
different for the highest-energy modes. There is a weaker drop of the
curve on the right towards zero and the top-most modes are badly
represented by the eigenmodes of~$C$. Thus there seem to be some
non-universal features in the manner that eigenmodes of~$C$ are
represented in the modes of~$A$; only the lowest modes are faithfully
represented in all systems. The representation of the topmost modes is
clearly model dependent.

\begin{figure}[tb]
  \centering
  \includegraphics[width=0.45\textwidth]{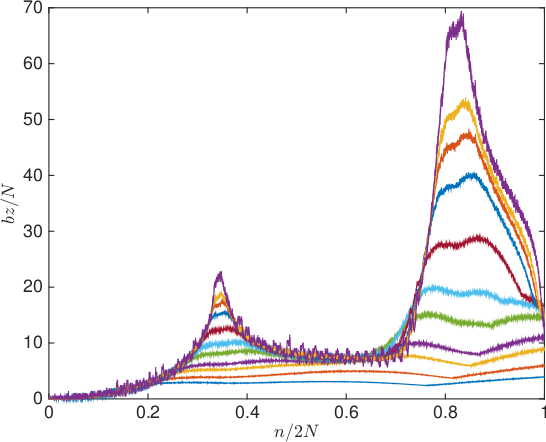}%
  \caption{ The 90\% bandwidth is plotted for the matrix~$W$ for a
    perfect crystal. $N=58^2$, for different~$z$ varying from $z=1$
    (flattest curves) to $z=3200$ (strongest peaks). We see that large
    numbers of modes are well reproduced for small~$n$ but far fewer
    modes are well reproduced at the top of the spectrum. The peaks
    are associated with the appearance of van Hove peaks in the
    density of states which seem to favour mode mixing, and a broader
    band in~$W$.}\label{fig:converg}%
\end{figure}

We plot in Fig.~\ref{fig:converg} the full bandwidth curves for a
crystal for $N=58^2$ for several different values of~$z$; In this
curve different parts of the spectrum behave in different ways. For
low energies and for $n/2N\sim 0.5$ there is a convergence of the
scaled curves for large~$z$. This is the same convergence behaviour that we saw
in Fig.~\ref{fig:band}. Very differently, for positions in the spectrum which seem
to be associated with the van Hove singularities there is a continuous
evolution of the spectrum with~$z$. This continuous evolution is {\it
  not seen for disordered systems} for which the whole curve seems to
stabilise for large~$z$ (data not shown).

\section{How many modes are reliable?}
\begin{figure}[tb]
  \centering
  \includegraphics[width=0.45\textwidth]{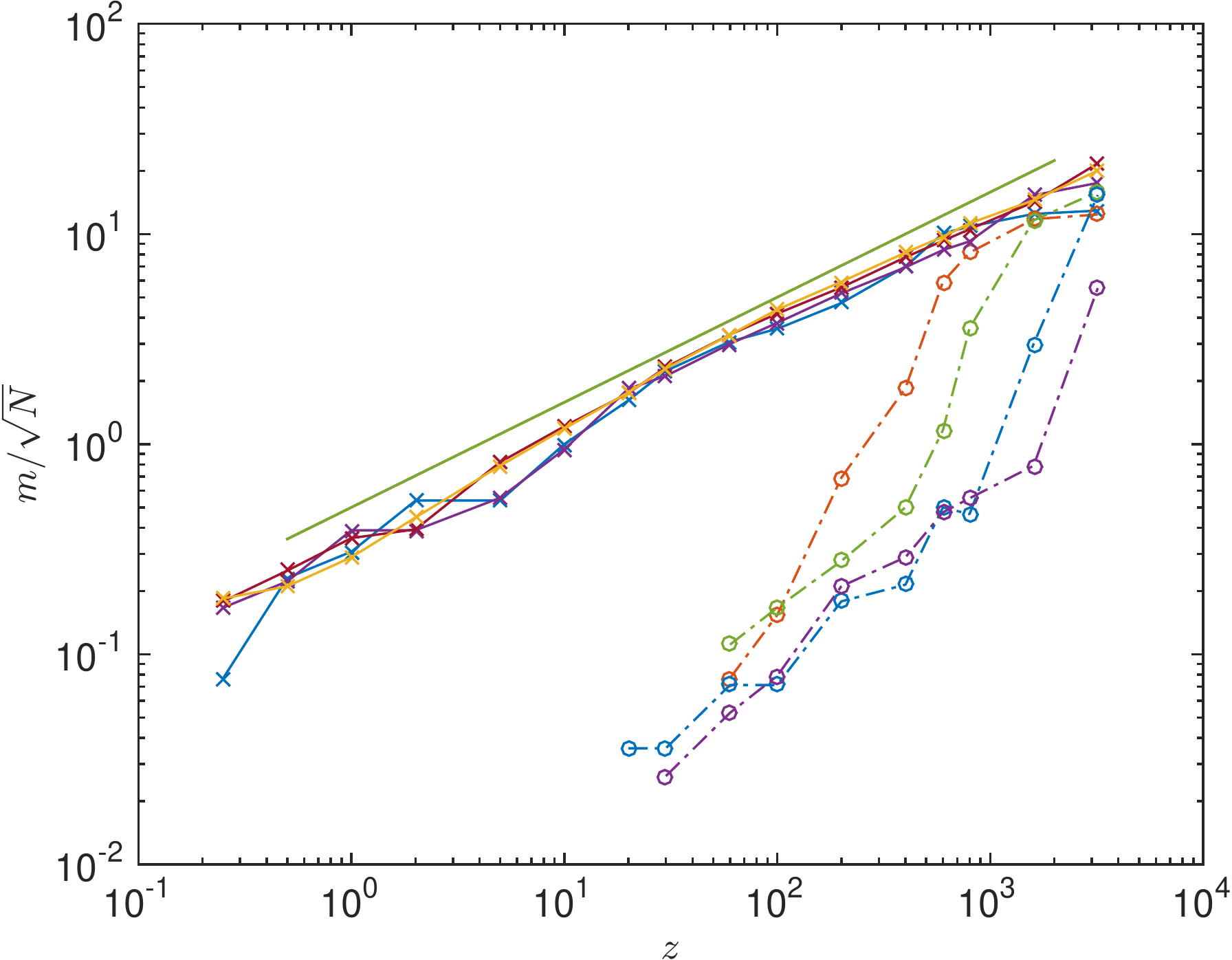}%
  \caption{ Number of good modes as a function of~$z$.  $N=18^2$,
    $N=28^2$, $N=38^2$, $N=58^2$. Straight line as a guide to the eye,
    $m=\sqrt{zN}/2$. Data from bottom of spectrum of~$A$, collapse to
    a single master curve.  Data from top of the spectrum do not
    collapse with this scaling. We only start seeing well resolved
    modes at the top of the spectrum for~$z>20$.  Numbers resolved at
    the top and bottom of the spectrum only become comparable for
    $z\sim1000$. Disordered system.}\label{fig:real}%
\end{figure}

We finish with a study of how many modes in~$C$ are reliable.
At both the bottom and the top of the spectrum of~$A$ we
find how many modes are reproduced within~$C$ with a matrix
element\footnote{In practice this was a diagonal element in the
  cases we visually checked as in Fig.~\ref{fig:overlap2}}
$W_{ij} >0.5$. We plot the results independently for the top and
bottom of the spectrum in Fig.~\ref{fig:real}. The most remarkable
result is a rather good empirical scaling so that the number of well
reproduced modes~$m$ at the bottom of the spectrum
\begin{equation}
  m \approx  \sqrt{Nz}/3 \sim \sqrt{T}\quad \text{ independent of N}
\end{equation}
This law works over large variations of $z$, $N$ and depends weakly on
the degree of disorder in the elastic medium.

The data coming from the top of the spectrum is much noisier and does
not exhibit a clean scaling with $N$ or~$z$. Indeed we also find
qualitatively different results between ordered and disordered systems
-- in an ordered system no modes are resolved at the top of the
spectrum. The resolution that we find in the disordered system is
perhaps linked to the localised nature of the modes.

\section{Effects of sampling rate}
We now consider the effects of finite relaxation times in an
experimental system. We know that long-wavelength modes relax more
slowly than those describing the shortest length scales. Since it is
these modes which are best described in the spectrum of~$C$ we might
fear a degradation of the method due to rapid sampling.
A full account would require a two fluid theory of
colloidal dynamics~\cite{hurd}, we here use a simpler description
with a Langevin equation which gives a qualitatively correct
description of the slowest, over-damped, longitudinal modes. Thus we
study a set of $2N$~coordinates evolving according to the equation
\begin{equation}
  \frac{d r}{ dt} = -A r +\xi(t) \label{eq:langevin}
\end{equation}
where $r$ is a vector and $\xi$ is a vector of Brownian noise,
$\langle\xi_i(t)\xi_j(s)\rangle = \delta_{ij}\delta(t{-}s)\,2/\beta$.

We sample the positions regularly with a time step~$\tau$ which is a fraction
$f$ of the relaxation rate of the slowest mode of the Langevin
equation.
\begin{equation}
  \tau= f/\lambda_1 \label{eq:f}
\end{equation}
with $\lambda_1$ the smallest eigenvalue of~$A$. This gives the
following update rule for the positions:
\begin{equation}
  r' = e^{-\tau A} r +\sum_i  {\mathcal N}(0, \sigma_i) v_i \label{eq:step}
\end{equation}
with ${\mathcal N}(0 ,\sigma)$ normally distributed random numbers
with mean zero and variance~$\sigma^2$. They are independent for different~$i$.
As above, $v_i$ is the eigenvector
corresponding to~$\lambda_i$.  For the equation~\eqref{eq:langevin} we find the variance for each mode,
\begin{equation}
  \sigma_i^2 = \frac{1}{\beta \lambda_i} \left[1 - e^{-2 \tau \lambda_i}  \right].
\end{equation}
We then build up our estimate of the covariance matrix using
eq.~\eqref{eq:corr} with $T=2Nz$ samples. When $\tau$ is large each
mode is sampled independently as in the Wishart ensemble considered
above. When $\tau$ is very small the positions remain highly
correlated between successive samples.
\begin{figure}[tb]
  \centering
  \includegraphics[width=0.45\textwidth]{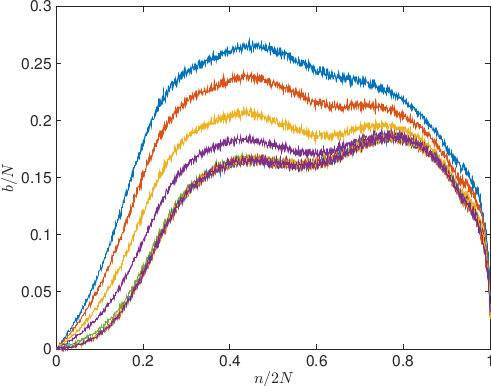}%
  \caption{ Effect of autocorrelation on the mode resolution:
    eq.~\eqref{eq:f}
    $f=[ .004\, .005\, .007\, .01\, .02\, .04\, .06\, .1\, .5\, 1\,
    2.]$
    from highest to lowest curve, $N=28^2$, $z=40$. Already curves
    with $f=.02$ are well converged. All curves for high values of $f$
    superpose. }\label{fig:frac}%
\end{figure}

We plot the bandwidth analysis of the dynamics, eq.~\eqref{eq:step}, in
Fig.~\ref{fig:frac}. We find the results most surprising: Already for the small
value $f=0.02$ the resolution of modes within~$C$ is very close to that found in
our above study of the Wishart ensemble. Even though the modes are sampled
very inefficiently, the diagonalisation is able to resolve the lowest modes
within the system. Indeed in the given example with $z=40$ and $f=0.02$ the
system is simulated for not quite one relaxation time of the slowest mode;
despite this the global appearance of the bandwidth is close to the fully
converged ensemble with the same value of~$z$.


\section{Conclusion}
We have carried out a numerical study of the mode structure of correlation
matrices, of the sort commonly extracted from colloidal materials. The
lowest eigenvectors of~$A$, (which correspond to the top eigenvectors
of~$C$) are rather easily extracted from the matrix. However within
the bulk of the spectrum there is a mixing of an extensive proportion
of the exact modes for values of~$z$ typically used in experiments. In
the case of disordered materials it is possible to study just a few
eigenvectors at the top of the spectrum of~$A$, even though the top
eigenvalues converge rather badly in the Marchenko-Pastur theory.
One of the most surprising features of our results is the convergence of
high-energy modes in the spectrum, which has not been observed in earlier
work~\cite{silke}. However the authors of this study worked very close to the
limit $z = 1$ where the convergence of these highest modes is not yet visible.
Such large-z studies have now been published by several experimental groups.

When we added the effect of finite relaxation times in the
construction of the correlation matrix we discovered that the lowest
modes are resolved with remarkably low statistics.


\end{document}